# Effects of Initial Attitude Estimation Errors on Loosely Coupled Smartphone GPS/IMU Integration System

Kwansik Park, Woohyun Kim, and Jiwon Seo*

School of Integrated Technology, Yonsei University,
Incheon, 21983, Korea (KwansikPark, crimy00, jiwon.seo@yonsei.ac.kr)

* Corresponding author

**Abstract**: Global Positioning System (GPS) and inertial measurement unit (IMU) sensors are commonly integrated using the extended Kalman filter (EKF), for achieving better navigation performance. However, because of nonlinearity, the performance of the EKF is affected by the initial state estimation errors, and the navigation solutions, including the attitude, diverge rapidly as the initial errors increase. This paper analyzes the data obtained from an outdoor experiment, and investigates the effect of the initial errors on the attitude estimation performance using EKF, which is used in loosely coupled low-cost smartphone GPS/IMU sensors.

## 1. INTRODUCTION

In various fields, the Global Positioning System (GPS) and inertial measurement unit (IMU) are used for navigation in numerous applications such as autonomous vehicles [1-4] and robots [5-8]. GPS is a global satellite-based navigation system that provides position, velocity, and timing solutions with long-term stability and zero accumulation of errors in time when four or more visible satellites are present [9-11]. However, the rate of GPS position output is relatively low, and the GPS receivers can undergo significant performance degradation under multipath, jamming, or ionospheric anomalies [12-16]. In addition, the attitude of a vehicle cannot be directly obtained using a single GPS antenna alone.

On the other hand, an IMU, with proper processing, can provide attitude as well as position and velocity outputs at relatively high rates. However, one of the major weaknesses of the IMU is that its errors accumulate rapidly as time progresses [17]. For these reasons, the integration of GPS and IMU has been studied extensively to complement and take advantage of their shortcomings and benefits [17].

To integrate the measurements from the two sensors (i.e., GPS and IMU), a loosely coupled integration method was adopted in [18], which utilized the extended Kalman filter (EKF) to estimate navigation solutions. However, one of the main disadvantages of the EKF was the divergence caused by the initial state estimate error [18].

In this study, the effect of the initial attitude estimation error on the loosely coupled GPS/IMU navigation system using low-cost sensors, implemented in a smartphone, is studied. An outdoor experiment is performed to collect data, and the deviations of the estimated roll, pitch, and yaw angles, according to the initial errors, are analyzed.

## 2. LOOSELY COUPLED GPS/IMU INTEGRATION WITH EXTENDED KALMAN FITER

This sensor integration system covered in this paper is described in Fig. 1.

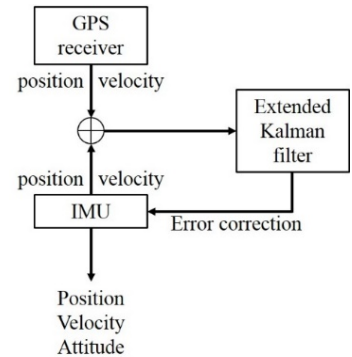

Fig. 1 Architecture of the loosely coupled GPS/IMU system using the EKF.

As shown in Fig. 1, both the GPS and IMU data are fused by the EKF that is used to estimate the errors of the IMU states. The estimated error states are fed back into the IMU for correction, and then, the corrected IMU outputs are used as the final navigation solution.

### 2.1 System and observation model

The discrete-time state-space model of the GPS/IMU integrated navigation system is expressed as follows [19]:

$$\delta \mathbf{x}_{k+1} = \mathbf{\Phi}_k \delta \mathbf{x}_k + \mathbf{G}_k \mathbf{u}_k + \mathbf{\zeta}_k \quad (1)$$

where $\delta \mathbf{x}_k$ is the error state vector at the $k$th time epoch, describing attitude (roll, pitch, and yaw), velocity, and position errors and biases of the gyroscopes and accelerometers. The matrices $\mathbf{\Phi}_k$ and $\mathbf{G}_k$ are the state-transition matrix and control-input matrix, respectively, and the vectors $\mathbf{u}_k$ and $\mathbf{\zeta}_k \sim N(\mathbf{0}, \mathbf{Q})$ are the sensor-measurement vector and process-noise vector that is assumed to has zero mean and covariance matrix $\mathbf{Q}$, respectively.

The observation model can be written as follows [19]:

$$\delta \mathbf{y}_k = \mathbf{H}_k \delta \mathbf{x}_k + \mathbf{v}_k \qquad (2)$$

where $\delta \mathbf{y}_k$ is the vector that represents the differences between the velocities and positions measured by the GPS and IMU, $\mathbf{H}_k$ is the observation matrix, and $\mathbf{v}_k \sim N(\mathbf{0}, \mathbf{R})$ is the measurement noise vector that is assumed to have zero mean and covariance matrix $\mathbf{R}$ [19]. The structures of all the matrices and vectors in (1) and (2) are given in detail in [19], and the noise matrices are typically obtained from the sensor specifications or calibrations [19].

### 2.2 EKF algorithm for the integration of GPS/IMU

After choosing the initial values of the state vector and covariance matrix, $\delta \hat{\mathbf{x}}_0$ and $\mathbf{P}_0$, the EKF predicts the state vector and covariance matrix of the next epoch as follows:

$$\delta \hat{\mathbf{x}}_k^- = \mathbf{\Phi}_k \delta \hat{\mathbf{x}}_{k-1} + \mathbf{G}_k \mathbf{u}_k$$

$$\mathbf{Q}_k = \mathbf{G}_k \mathbf{Q} \mathbf{G}_k^T \Delta T \qquad (3)$$

$$\mathbf{P}_k^- = \mathbf{\Phi}_k \mathbf{P}_{k-1} \mathbf{\Phi}_k^T + \mathbf{Q}_k$$

where $\delta \hat{\mathbf{x}}_k^-$ and $\mathbf{P}_k^-$ are the predicted state vector and covariance matrix, respectively, and $\Delta T$ is the time interval between two epochs. After the prediction, the Kalman gain, $\mathbf{K}_k$, and the estimated values of the state vector and covariance matrix, $\delta \hat{\mathbf{x}}_k$ and $\mathbf{P}_k$, are updated as follows:

$$\mathbf{K}_k = \mathbf{P}_k^- \mathbf{H}_k^T (\mathbf{H}_k \mathbf{P}_k^- \mathbf{H}_k^T + \mathbf{R})^{-1}$$

$$\delta \hat{\mathbf{x}}_k = \delta \hat{\mathbf{x}}_k^- + \mathbf{K}_k (\delta \mathbf{y}_k - \mathbf{H}_k \delta \hat{\mathbf{x}}_k^-) \qquad (4)$$

$$\mathbf{P}_k = (\mathbf{I} - \mathbf{K}_k \mathbf{H}_k) \mathbf{P}_k^-$$

The prediction and update steps in (3) and (4) are repeated.

When choosing the initial state vector for the GPS/IMU integrated navigation system, the initial attitude estimated at rest before moving along the trajectory is used. The EKF is not an optimal estimator, in general, and the estimation error of the initial state can cause the EKF to diverge [18]. Therefore, the estimation error of the initial attitude can cause deviations in the attitude estimation of the EKF. The effects of the initial estimation error are analyzed by an outdoor experiment, and are described in the next section.

### 3. EXPERIMENTAL RESULTS

In order to investigate the deviations in the attitude estimation of the EKF, due to the initial attitude estimation error, the actual measurement data from the GPS and IMU sensors of a Galaxy A5 smartphone were logged along the trajectory shown in Fig. 2. Before moving along the trajectory, two hours of data from the GPS and IMU were recorded and used for 1) analyzing the sensor error characteristics that are required to construct the noise covariance matrices $\mathbf{Q}$ and $\mathbf{R}$, and 2) estimating the initial attitude. The analyzed and acquired sensor parameters of the three gyroscopes and three accelerometers, corresponding to the orthogonal axes and the estimated initial attitude, are listed in Tables 1 through 3.

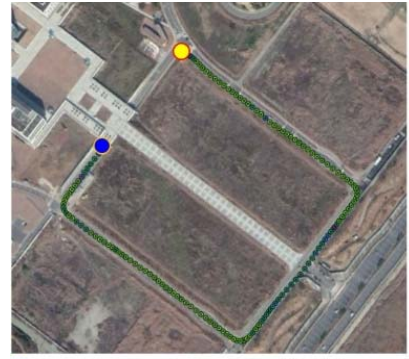

○ Start  ● Finish

Fig. 2 Trajectory of the experiment.

Table 1 Analyzed parameters of the gyroscopes from Galaxy A5 based on two-hour sensor data.

| Parameter | Value |
|---|---|
| Angle random walks | $\begin{bmatrix} 7.1242 \times 10^{-6} \\ 5.9828 \times 10^{-6} \\ 5.7239 \times 10^{-6} \end{bmatrix} (\frac{\text{rad/s}}{\sqrt{\text{Hz}}})$ |
| Static bias | $\begin{bmatrix} -5.7265 \times 10^{-6} \\ -5.2920 \times 10^{-6} \\ 5.2511 \times 10^{-6} \end{bmatrix}$ (rad/s) |
| Dynamic bias | $\begin{bmatrix} 2.0040 \times 10^{-7} \\ 1.9759 \times 10^{-7} \\ 2.1179 \times 10^{-7} \end{bmatrix}$ (rad/s) |
| Dynamic bias power spectral density (PSD) | $\begin{bmatrix} 3.1686 \times 10^{-6} \\ 3.1241 \times 10^{-6} \\ 3.3487 \times 10^{-6} \end{bmatrix} (\frac{\text{rad/s}}{\sqrt{\text{Hz}}})$ |
| Correlation time | $\begin{bmatrix} 1000 \\ 1000 \\ 1000 \end{bmatrix}$ (s) |

Table 2 Analyzed parameters of the accelerometers in Galaxy A5 based on two-hour sensor data.

| Parameter | Value |
|---|---|
| Velocity random walks | $\begin{bmatrix} 0.0013 \\ 0.0024 \\ 0.0030 \end{bmatrix} (\frac{\text{m/s}^2}{\sqrt{\text{Hz}}})$ |
| Static bias | $\begin{bmatrix} 0.1280 \\ 0.0095 \\ 9.7601 \end{bmatrix}$ (m/s$^2$) |
| Dynamic bias | $\begin{bmatrix} 0.0011 \\ 0.0018 \\ 0.0016 \end{bmatrix}$ (m/s$^2$) |
| Dynamic bias PSD | $\begin{bmatrix} 0.0030 \\ 0.0128 \\ 0.0135 \end{bmatrix} (\frac{\text{m/s}^2}{\sqrt{\text{Hz}}})$ |
| Correlation time | $\begin{bmatrix} 30 \\ 200 \\ 300 \end{bmatrix}$ (s) |

Table 3 Initial attitude estimation (radians).

|  | Roll | Pitch | Yaw |
|---|---|---|---|
| **Values** | −0.0068 | 0.0418 | 1.2234 |

Errors were added intentionally to the three angles in Table 3 to observe their effect. The three angular errors were set to be the same for simplicity. The value of the angular error, ε, was increased from 0° to 0.1°. Then, for each value of ε, the roll, pitch, and yaw angles were estimated from the EKF in (3) and (4) along the trajectory presented in Fig. 2. As a result, the root mean square (RMS) of the deviations between the attitude estimations based on the initial attitude with each value of ε and the original initial attitude in Table 3 are presented in Fig. 3. The RMS of each attitude was calculated after finishing the trajectory in Fig. 2.

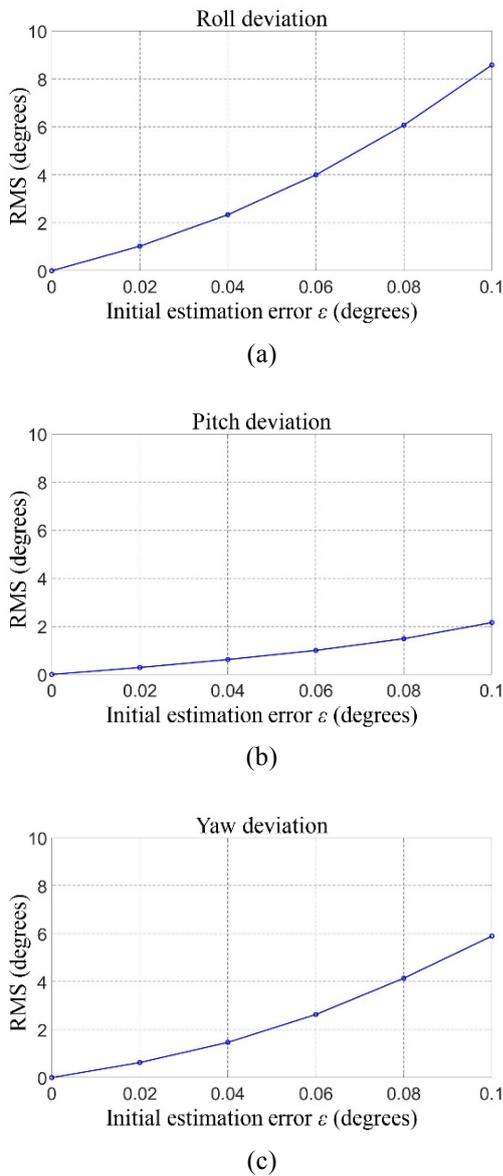

Fig. 3 RMS of (a) roll, (b) pitch, and (c) yaw deviations versus various values of initial estimation error $\varepsilon$.

As shown in Fig. 3, the estimated roll, pitch, and yaw angles tend to deviate rapidly as the initial estimation errors increase. The maximum deviations of the roll, pitch, and yaw angle are approximately 8.5°, 2°, and 6°, respectively, despite the relatively small initial error ($\varepsilon$ = 0.1°). These results show the sensitivity of the EKF to the initial estimation error. Note that the integrated system needs to achieve the RMS attitude errors within 1° [20] for realistic applications such as unmanned aerial vehicle.

## 4. CONCLUSIONS

This paper presented the effects of the initial error on the attitude estimation of the EKF for loosely coupled GPS/IMU integration using low-cost GPS and IMU sensors in a smart phone. Real sensor data were collected from an outdoor experiment, and used to analyze the deviations in the attitude estimations from the EKF, according to the initial estimation errors. The experimental results showed that the estimated roll, pitch, and yaw angles deviated rapidly from the original values as the initial estimation errors increased. Thus, it is necessary to develop a method that can provide robustness against the effect of the initial estimation errors, for achieving better navigation performance using low-cost sensors.

## ACKNOWLEDGEMENT

This research was supported by the Unmanned Vehicles Core Technology Research and Development Program through the National Research Foundation of Korea (NRF) and the Unmanned Vehicle Advanced Research Center (UVARC) funded by the Ministry of Science and ICT, Republic of Korea (No. 2020M3C1C1A01086407).